\documentclass[12pt,draftcls,onecolumn]{IEEEtran}
\date{}
\long\def\symbolfootnote[#1]#2{\begingroup\def\thefootnote{\fnsymbol{footnote}}
\footnote[#1]{#2}\endgroup}

\usepackage{cite}
\usepackage{url}

\usepackage{graphicx}
\graphicspath{{pix/}}
\usepackage{subfigure}
\usepackage{setspace}

\usepackage{algorithm}
\usepackage{algorithmic}
\usepackage{amsmath, amssymb}

\usepackage{balance}
\usepackage{capt-of}

\begin{document}

\title{Security of Spectrum Learning in Cognitive Radios}

\author{\IEEEauthorblockN{Behnam Bahrak and Jung-Min ``Jerry''  Park}\\
\IEEEauthorblockA{Department of Electrical and Computer Engineering\\
Virginia Tech, Blacksburg, VA, 24061\\
Email: \{bahrak, jungmin\}@vt.edu}
}

\maketitle

\begin{abstract}
Due to delay and energy constraints, a \emph{cognitive radio} may not be able to perform spectrum sensing in all available channels. Therefore, a sensing policy is needed to decide which channels to sense. The \emph{channel selection} problem is the problem of designing such a sensing policy to maximize throughput while avoiding interference to primary users. The channel selection problem can be formulated as a reinforcement learning problem. Channel selection schemes that employ reinforcement machine learning algorithms are vulnerable to \emph{belief manipulation attacks} that contaminate the knowledge base of the learning algorithms. In this paper, we analyze the security of channel selection algorithms that are based on reinforcement learning and propose mitigation techniques that make these algorithms more robust against belief manipulation attacks.
\end{abstract}


\section{Introduction}\label{sec:intro}

It is widely believed that cognitive radios (CRs) are one of the key technologies that can address the spectrum scarcity problem. It is expected that they will play an important role in maximizing spectrum utilization and help satisfy the QoS requirements of a number of important communications applications---from emergency first responders' public safety networks to military tactical networks. CRs often employ software-defined radio platforms that are capable of executing complex computational tasks to communicate efficiently without causing interference to licensed (a.k.a. primary) users. A specialized software module within a CR called the \emph{cognitive engine} performs the aforementioned tasks, such as the optimization of the transmission/reception (TX/RX) parameters and execution of spectrum sensing and spectrum access strategies. Most of the tasks performed by a cognitive engine require the use of machine learning algorithms, especially if those tasks need to be carried out in a distributed manner. 

Considering the computing limitations and energy constraints of a battery-powered CR, a CR may not be able to perform full-spectrum sensing (i.e., sense all available spectrum bands) because of its prohibitive cost. Therefore, a spectrum sensing policy at the medium access control (MAC) layer is needed to decide which set of channels to sense. The \emph{channel selection problem} is the problem of designing such a sensing policy. The optimal channel selection strategy for an unlicensed user (i.e., secondary user) is based on the availability statistics of the channels. The availability of the channels is determined by the presence/absence of primary user signals in those channels. The channels' availability statistics are initially unknown to a secondary user and need to be estimated using sensing samples. The critical tradeoff that the cognitive engine faces in each timeslot is between transmission (``exploitation") on the channel that has the highest expected reward (e.g., throughput) and channel sensing (``exploration") to get more information about the expected rewards of the other channels. The exploitation vs. exploration tradeoff problem, such as the one just described, is central to an area of machine learning known as \emph{reinforcement learning}.

\emph{Spectrum learning} is the process of learning the spectrum statistics (i.e., primary user occupancy information), which is crucial to enable CRs to sense/interpret their spectrum environment and make intelligent decisions to achieve efficient communication. Although spectrum learning is beneficial for CRs, it can pose a serious security vulnerability. A radio that can learn has the potential to be taught by malicious entities in an adversarial environment. This kind of threat may have a long-lasting impact on the cognitive radio network.

As mentioned above, the design of the optimal sensing policy can be formulated as a \emph{reinforcement learning} (RL) problem. When the channels are assumed to be independent, it can be formulated as a special class of RL problems known as a \emph{restless multi-armed bandit} process. Recent results (i.e., \cite{bahrak:en:czs:08} -- \cite{bahrak:en:al:09}) show that a surprisingly simple \emph{myopic policy} that ignores the impact of the current action on the future reward is optimal when channels are identical. In this paper, we show that in adversarial environments, where an active attacker performs belief manipulation attacks against the machine learning algorithms executed on a cognitive engine, the myopic policy is no longer optimal and a \emph{softmax policy} that exploits some level of randomness outperforms the myopic policy.

Our contributions can be summarized as follows:
\begin{enumerate}
\item We derive closed-form expressions for the throughput of cognitive radios in an adversarial environment for the two-channel case in two channel selection policies, viz myopic policy and softmax policy.
\item We derive the attacker's optimal attack strategy and the cognitive engine's optimal defense strategy by solving respective optimization problems for more than two non-identical channels.
\item We identify and discuss two fundamental trade-offs in the security of spectrum learning: (1) the attackerÕs tradeoff between the attack probability and the number of required observations for attack detection (i.e., the time that the attack detection system needs to detect an attack); and (2) the channel selection systemÕs tradeoff between attack resilience and performance of a given channel selection policy.
\item We prove that for sufficiently large attack probabilities, a softmax policy with a proper choice of parameters outperforms the myopic policy for all possible attacker's strategies. 
\end{enumerate}

The rest of this paper is organized as follows. In Section~\ref{sec:related}, the work related to this paper are discussed. Section~\ref{sec:model} provides the channel selection system model and the attack model. In Section~\ref{sec:two}, we analyze sensing policies in an adversarial environment for a cognitive radio system with two channels and Section~\ref{sec:more} extends the result to a cognitive radio system with more than two channels. Finally Section~\ref{sec:con} concludes the paper.

\section{Related Work}\label{sec:related}

The formulation of spectrum learning problem as a restless multi-armed bandit process is investigated in \cite{bahrak:en:czs:08}--\cite{bahrak:en:lejp:11}. It is proved that when channels are identical and independent the myopic policy is the optimal policy \cite{bahrak:en:zkl:08}--\cite{bahrak:en:al:09}, and that this policy is a special case of Whittle's index policy for the restless bandit problem which can be computed for non-identical channels as well \cite{bahrak:en:lz:08}. An asymptotically optimal policy is proposed in \cite{bahrak:en:lejp:11} for a more realistic case where the policy does not require any prior statistical knowledge about the traffic pattern and the channels are different. Despite all these work that assume a non-adversarial environment, this paper investigates the spectrum learning problem in an adversarial environment where active attackers aim to reduce the throughput of the cognitive radio network.

The security of machine learning algorithms that have been applied to applications such as intrusion detection systems (IDS) and spam filters is investigated thoroughly in \cite{bahrak:en:lm:05}-\cite{bahrak:en:tgmps:11}. In these papers, the authors discuss how an adversary can maliciously mistrain a learning system in an IDS and how an attacker may contaminate the knowledge base of a spam email filtering system to bypass the filtering. In \cite{bahrak:en:bns:06} and \cite{bahrak:en:bnj:08} different kinds of attacks against machine learning algorithms are introduced and a variety of potential defenses against those attacks are proposed. 

In the context of cognitive radios, security of machine learning algorithms that are used for signal classification are addressed in \cite{bahrak:en:nc:09} and \cite{bahrak:en:ckn:11}, but the types of learning algorithms that are analyzed is different from the algorithms in this paper. To the best of our knowledge, this paper describes the first analysis of a reinforcement learning algorithm's vulnerability against belief manipulation attacks to cognitive radios.

\section{The Channel Selection System and the Attack Model}\label{sec:model}
In this section, we introduce the channel selection system and establish the attack model. 

\subsection{Channel Selection System Model}
We consider a general dynamic spectrum access system where a user has access to $N$ independent and stochastically non-identical parallel Gillbert-Elliot channels \cite{bahrak:en:g:60}, and chooses one channel to sense and access in each time slot, aiming to maximize its expected long-term reward (i.e., throughput). 

\begin{figure}
\centering 
    \includegraphics[width=4in]{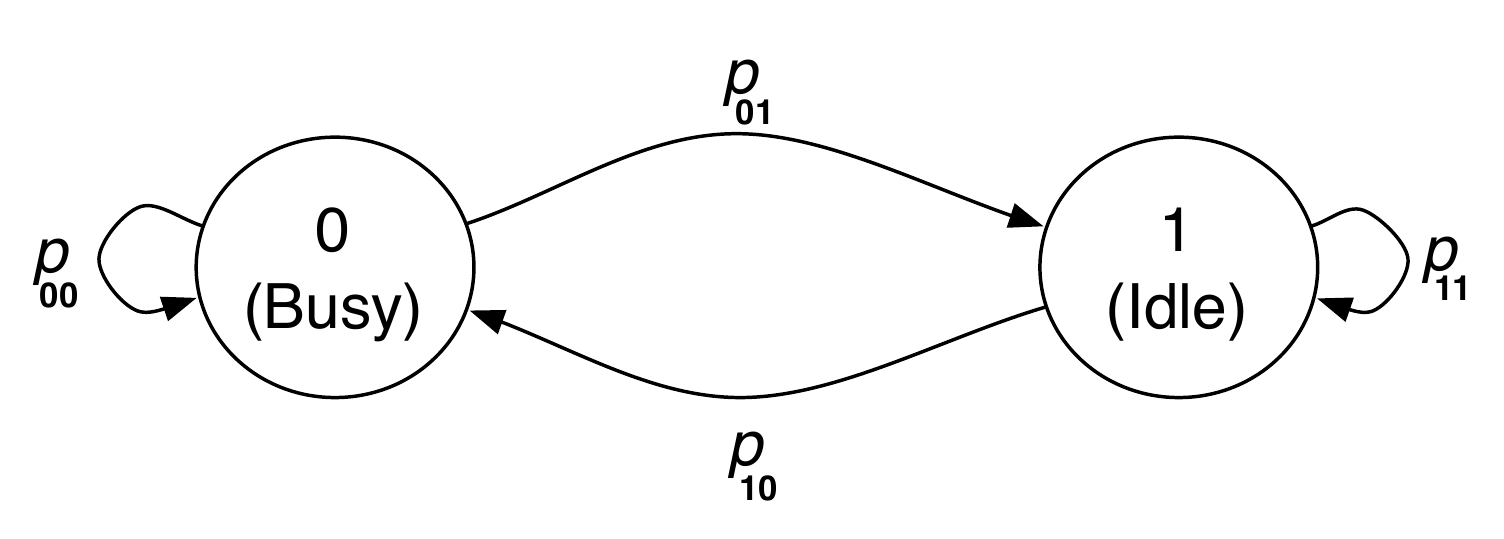}
\caption{A Gillbert-Elliot channel.}
\label{fig:gil}
\end{figure}

As illustrated in Figure~1, the state of the $k$-th channel---either idle (1) or busy (0)---indicates that the channel is unused by primary users or it is occupied. The transitions between these two states follow a Markov chain with transition probabilities $\{ p^{k}_{ij} \}_{i,j = 0,1}$. We assume that these transition probabilities for all channels are learnt in a non-adversarial environment before the system starts operating and thus the transition probabilities are known to the system. We also assume that $p^{k}_{11} > p^{k}_{01}$ or equivalently the channel states in two consecutive time slots are positively correlated. Note that this assumption is only used for derivation of closed-form expressions and can easily be relaxed by separately considering the case where this assumption does not hold. Due to its limited sensing and access capability, a secondary user chooses one of the $N$ channels to sense and access in each slot. Designing an optimal sensing policy that governs the channel selection at each time slot can be formulated as a restless multi-armed bandit process for independent channels. We denote $S_{k}(t)$ as the state of channel $k$ in slot $t$ that is given by the two-state Markov chain in Figure~\ref{fig:gil}. Let $S(t) = [S_{1}(t), \cdots ,S_{N}(t)] \in \{0,1\}^{N}$ denote the full system state.

The channel selection system is reward-based. At each time slot the secondary user selects one of the $N$ channels to sense. If the sensed channel is occupied by primary user signals, the user collects no reward; otherwise it accesses the channel and collects one unit of reward. The system keeps periodical sensing and transmitting on that channel until a primary user appears on the channel or a jamming attack prevents the user from transmission on the channel.  The secondary user's aim is to maximize the \emph{throughput} (reward) over a horizon of $T$ slots by choosing an optimal sensing policy.

Due to limited sensing (i.e., sensing only one channel out of $N$ channels), the full system state ($S(t)$) in slot $t$ is not observable. However, it has been shown that a sufficient statistic for optimal decision making is given by the conditional probability that each channel is in state $1$, given all past observations and decisions \cite{bahrak:en:ss:71}. Referred to as the belief vector, we denote this sufficient statistic by $\Omega(t) = [\omega_{1} , \cdots , \omega_{N}]$, where $\omega_{k}(t)$ is called the belief value of channel $k$ which is equivalent to the conditional probability that $S_{k}(t) = 1$ given all past observations and decisions for that channel. Given the sensing action $a(t) = k$ (the channel that is selected to be sensed in slot $t$) and the observation $S_{k}(t)$ in slot $t$, the belief vector for slot $t+1$ can be updated via Bayes rule through the following equation:
\begin{equation}\label{update}
\omega_{k}(t+1)= \left \{ \begin{array}{ll} p^{k}_{11}, & a(t)=k , S_{k}(t) = 1  \\ p^{k}_{01},  & a(t) = k , S_{k}(t) = 0 \\ \Gamma(\omega_{k}(t)), & a(t) \neq k \end{array} \right.,
\end{equation}
where $\Gamma(x) = xp^{k}_{11} + (1-x)p^{k}_{01}$.

A sensing policy $\pi$ specifies a sequence of functions $\pi = [\pi_{1} , \cdots , \pi_{t}]$, where $\pi_{t}$ maps the belief vector $\Omega(t)$ to a sensing action $a(t)$. Multi-channel opportunistic access can thus be formulated as the following stochastic optimization problem:
\begin{equation}\label{stoc}
\pi^{*} = \arg \max_{\pi} E_{\pi} [\sum_{t=1}^{T} R_{\pi_{t}(\Omega(t))}(t)|\Omega(1)],
\end{equation}
where $\pi_{t}(\Omega(t))$ is the channel selected for sensing and $R_{\pi_{t}(\Omega(t))}(t)$ is the reward when the belief vector is $\Omega(t)$ and the action $\pi_{t}(\Omega(t))$ is taken, and $\Omega(1)$ is the initial belief vector. If no information about the initial system state is available, each entry of $\Omega(1)$ can be set to the stationary distribution $\omega^{k}_{0}$ of the underlying Markov chain: 
\begin{equation}\label{omeg}
\omega^{k}_{0} = \frac{p^{k}_{01}}{p^{k}_{01}+p^{k}_{10}}.
\end{equation}

Let $V_{t}(\Omega(t))$ be the value function which represents the maximum expected total reward that can be obtained starting from slot $t$ given the current belief vector $\Omega(t)$. Given that the user selects channel $k$ and observes $S_{k}(t)$ in slot $t$, the maximum expected reward consists of the following two parts:
\begin{enumerate}
\item The expected immediate reward: \begin{equation*} E[R_{k}(t)] = E[S_{k}(t)] = \omega_{k}(t). \end{equation*}
\item The maximum expected future reward:  \begin{equation*} V_{t+1}(\tau(\Omega(t) | k,S_{k}(t))),\end{equation*}  where $\tau(\Omega(t)|k , S_{k}(t))$ denotes the updated belief vector for slot $t+1$ as given in (1). If we maximize over all channel selections, we obtain the following optimization equation:
\begin{equation*}\label{reward}
V_{t}(\Omega(t)) = \max_{k=1,\cdots ,N} \{ \omega_{k}(t) + V_{t+1}(\tau(\Omega(t)|k , S_{k}(t))) \}.
\end{equation*}
\end{enumerate}
Because the value function is limited to horizon $T$, we have: $V_{T}(\Omega(T)) = \max_{k = 1 , \cdots , N} \omega_{k}(T)$ and $V_{t}(\Omega(t))= 0$ for $t>T$.
Theoretically, the optimal policy $\pi^{*}$ and its performance $V_{1}(\Omega(1))$ can be obtained by solving the above dynamic programming problem. However, because of the impact of the current action on the future reward and the uncountable space of the belief vector, obtaining the optimal solution via the above recursive equations is computationally prohibitive. 

\subsection{Attack Model}
We assume that there exists a single attacker in the environment who tries to decrease the throughput of secondary users by preventing them from transmission in some time slots, by employing adaptive interference techniques (or cognitive jamming attacks). Although many defenses against these attacks have been proposed, but none of these defenses is perfect or can address all classes of attackers. While such defenses can restrict the instantaneous effect of these attacks, they can not reduce the long-term effect of such attacks on cognitive radio network, when the network is using a learning system as part of its cognitive engine. These attacks gradually contaminate the knowledge base of a cognitive radio and lead the learning system to make wrong decisions which results in degrading the performance of the radio.

The attacker uses the same equipment as secondary users, i.e., it is equipped with a CR that is comparable to a typical secondary user's CR in terms of battery capacity, transmission power, computing power, memory capacity, etc. Therefore the attacker is power-limited and wish to avoid jamming continuously, which quickly drains power and causes fast detection by the attack detection module of cognitive engine. We also assume that the attacker can attack only one channel at each time slot. Suppose that the attacker perform attacks in $t_{a}$ slots out of $T$ consecutive time slots on average. We denote $\alpha = \frac{t_{a}}{T}$ as the \emph{attack probability} in the environment. The attacker controls the attack probability in order to cause maximal damage to the network in terms of throughput reduction. It can easily be seen that a higher attack probability will result in a lower throughput for the cognitive radio system.

We assume that the adversary possesses full knowledge about the network and its parameters. We introduce the notion of the attacker's optimal strategy using the following definition:

 \newtheorem{D0}[definition]{Definition}
\begin{D0}
The attacker's {\bf $\alpha$-optimal strategy} is a strategy that minimizes the throughput of the target cognitive radio while keeping the attack probability fixed to value $\alpha$.
\end{D0}

\subsection{Attack Detection Model}

The cognitive radio network employs a mechanism for monitoring network status and detecting potential malicious activity. This monitoring mechanism is proposed in \cite{bahrak:en:lkp:10} to protect the network against jamming attacks. The monitoring can be done by specific monitor nodes in a distributed network and a detection algorithm is employed by the detection module at a monitor node; it takes as input observation samples obtained by the monitor node (i.e., failed transmission/successful transmission) and decides whether there is an attack or not. On one hand the observation window should be small enough, such that the attack is detected in a timely manner and appropriate countermeasures are initiated. On the other hand, this window should be sufficiently large such that the chance of a false alarm or a mis-detection is reduced.

The sequential nature of observations at consecutive time slots motivates the use of sequential detection techniques. A sequential decision rule is efficient if it can provide reliable decisions as fast as possible. There exists a trade-off between detection delay and detection accuracy in a detection scheme, i.e. a faster decision unavoidably leads to higher values of the probability of false alarm $P_{FA}$ and probability of mis-detection $P_{M}$ while lower values of these probabilities are attained at the expense of detection delay. For given values of $P_{FA}$ and $P_{M}$, the detection test that minimizes the average number of required observations (and thus average delay) to reach a decision among all sequential and non-sequential tests is \emph{Wald's Sequential Probability Ratio Test (SPRT)} \cite{bahrak:en:w:47}. 

In our case, the test is between hypotheses $H_{0}$ and $H_{1}$ with Bernoulli probability mass functions (p.m.fs) $f_{0}$ and $f_{1}$. Assume that $Y$ is a random variable with the Bernoulli distribution, where $Y=1$ denotes a failed transmission event in a slot. $H_{0}$ denotes the hypothesis that assumes the absence of an attack, and because the probability of a failed transmission in the absence of attack is $p_{10}$ (i.e. $Pr\{Y=1\} = p_{10}$), the corresponding p.m.f $f_{0}$ has a Bernoulli distribution with parameter $\theta_{0} = p_{10}$. Similarly, because the probability of a failed transmission in the presence of an attack with attack probability $\alpha$ is $1 - p_{11}(1-\alpha)$ (i.e. $Pr\{Y=1\} = 1 - p_{11}(1-\alpha)$), $H_{1}$ that denotes the hypothesis that assumes the existence of an attack has a Bernoulli p.m.f $f_{1}$ with parameter $\theta_{1} = 1 - p_{11}(1-\alpha)$.

The logarithm of likelihood ratio at stage $k$ with observations $x_{1}, \cdots ,x_{k}$ is:
\begin{equation*}
S_{k} = \sum_{i=1}^{k} \ln \frac{f_{1}(x_{i})}{f_{0}(x_{i})}.
\end{equation*}
where $x_{i} = 1$, if the system observes a failed transmission at time slot $i$, and $x_{i} = 0$ otherwise.  

The decision variable is defined as follows: 
\begin{equation*}
S_{k} \geq a  \Rightarrow \mbox{ accept } H_{1} 
\end{equation*}
\begin{equation*}
S_{k} < b  \Rightarrow \mbox{ accept } H_{0} 
\end{equation*}
\begin{equation*}
b \leq S_{k} < a  \Rightarrow \mbox{ take another observation}
\end{equation*}

The analysis in \cite{bahrak:en:lkp:10} shows that using this SPRT, the average number of samples needed for detecting an attack is:
\begin{equation*}\label{ASN}
E[N|H_{1}] = \frac{C}{\theta_{1} \ln(\frac{\theta_{1}}{\theta_{0}}) + (1-\theta_{1}) \ln(\frac{1-\theta_{1}}{1-\theta_{0}})},
\end{equation*}
where $C$ is a fixed positive number.
Using this equation for $E[N|H_{1}]$ we have:
\begin{align*}
\frac{\partial E[N|H_{1}]}{\partial \alpha} &= \frac{\partial E[N|H_{1}]}{\partial \theta_{1}} \times \frac{\partial \theta_{1}}{\partial \alpha} \\
&= \frac{-C(\ln(\frac{\theta_{1}}{\theta_{0}}) + \ln(\frac{1-\theta_{0}}{1-\theta_{1}}))}{(\theta_{1} \ln(\frac{\theta_{1}}{\theta_{0}}) + (1-\theta_{1}) \ln(\frac{1-\theta_{1}}{1-\theta_{0}}))^{2}} \times p_{11}.
\end{align*}

Because $\theta_{1} > \theta_{0}$, we have $\frac{\partial E[N|H_{1}]}{\partial \alpha} < 0$ and consequently $E[N|H_{1}]$ is a decreasing function of $\alpha$, i.e., increasing the attack probability would decrease the average number of required observations for attack detection. This poses a fundamental trade-off problem to an attacker: Increasing the attack probability, $\alpha$, increases the impact of the attack on the target (i.e., lower its throughput), but it also enables a detection module to detect the attack sooner. We define the attacker's cost as the inverse of $E[N|H_{1}]$:
\begin{equation*}
 \frac{\theta_{1} \ln(\frac{\theta_{1}}{\theta_{0}}) + (1-\theta_{1}) \ln(\frac{1-\theta_{1}}{1-\theta_{0}})}{C},
\end{equation*}
 Also in order to normalize the attacker's cost, we assume that the constant $C$ is equal to the maximum value of the statement $\theta_{1} \ln(\frac{\theta_{1}}{\theta_{0}}) + (1-\theta_{1}) \ln(\frac{1-\theta_{1}}{1-\theta_{0}})$ that happens at $\alpha = 1$, i.e. $C = \ln(\frac{1}{p_{10}})$, therefore:
 \begin{equation}\label{attackcost}
 \mbox{Attacker Cost} = \frac{\theta_{1} \ln(\frac{\theta_{1}}{\theta_{0}}) + (1-\theta_{1}) \ln(\frac{1-\theta_{1}}{1-\theta_{0}})}{\ln(\frac{1}{p_{10}})}.
 \end{equation}
 This definition for the attacker's cost shows that by risking detection of the attack by a detection module, the attacker's cost increases. The equation~\ref{attackcost} would be used as a measure for the \emph{attacker's cost} in the rest of this paper.

\section{Sensing Policies Analysis in an Adversarial Environment with two channels}\label{sec:two}
In this section, we analyze and compare the myopic policy \cite{bahrak:en:zkl:08} and the softmax policy \cite{bahrak:en:sb:98} in a hostile environment for $N=2$ identical channels, i.e. $p^{1}_{ij} = p^{2}_{ij} = p_{ij}$. 

\subsection{Analysis of the Myopic Policy}\label{ssec:2myopic}
The myopic policy explained in this section is identical to the one presented in \cite{bahrak:en:zkl:08}, but in this section, we analyze the performance of this policy in an adversarial environment.
A myopic policy ignores the impact of the current action on the future reward and only focuses on maximizing the immediate reward. At any given  time slot $t$, the myopic policy for selecting a channel for sensing can be expressed as follows:
\begin{equation*}
\pi^{m}(t) = \arg \max_{i = 1, \cdots , N} \omega_{i}(t).
\end{equation*}

In \cite{bahrak:en:alj:09}, the authors proved that for the channel selection system that was introduced in Section~2.1, the myopic policy is the optimal policy for all $N$. They also showed that the myopic policy has a simple structure that does not require the knowledge of the transition probabilities $p_{ij}$ or updates to the belief vector.

We define the steady-state throughput of the myopic policy as:
\begin{equation*}
U^{m} = \lim_{T \rightarrow \infty} \frac{V^{m}_{1:T}(\Omega(1))}{T},
\end{equation*}
where $V^{m}_{1:T}(\Omega(1))$ is the expected total reward obtained in $T$ slots under the myopic policy when the initial belief vector is $\Omega(1)$. The key to computing the throughput $U$ is to first find how long a user stays in the same channel. Let us introduce the concept of a \emph{transmission period} (TP), which represents the time that a user stays in the same channel. Let $L_{k}$ denote the $k$-th TP. In \cite{bahrak:en:alj:09}, Zhao et al. showed that under the condition $p_{11} > p_{01}$, the steady-state throughput is:
\begin{equation}\label{thr}
U = 1 - \frac{1}{\overline{L}},
\end{equation}
where $\overline{L} = \lim_{K \rightarrow \infty} \frac{\sum_{k=1}^{K}L_{k}}{K}$ denotes the average length of a TP. 

Throughput analysis is thus reduced to analyzing the average TP length $\overline{L}$. For $N=2$, we can derive a closed-form expression of $\overline{L}$ as a function of the attack probability $\alpha$, which leads to a closed-form expression of the myopic policy throughput $U^{m}(\alpha)$.

 \newtheorem{T1}[theorem]{Theorem}
 \begin{T1}
For $N=2$, the average TP length of a myopic policy as a function of the attack probability, $\alpha$, is given by:
\begin{equation}\label{T1L}
L^{m}(\alpha) = 1 + \frac{\overline{\omega}}{1-p_{11}(1-\alpha)},
\end{equation}
where 
\begin{equation}\label{T1O}
\overline{\omega} = \frac{(1 - \alpha)p_{01}^{(2)}}{(1 - \alpha)p_{01}^{(2)} - A},
\end{equation} and
 \begin{equation}\label{T1A}
 A = \omega_{0} (1 - \alpha) [1 - \frac{(p_{11}-p_{01})^{3}(1 - p_{11}(1-\alpha))}{1 - p_{11}(1-\alpha)(p_{11} - p_{01})}],
\end{equation}
and 
\begin{equation*}
p_{01}^{(2)} = \frac{p_{01} - p_{01}(p_{11}-p_{01})^{2}}{p_{01}+p_{10}}.
\end{equation*}
\end{T1}

 \begin{proof}
From the structure of the myopic policy, $\{ L_{k} \}_{k=1}^{\infty}$ forms a first-order Markov chain for N = 2. When the system is running in an adversarial environment with attack probability $\alpha$, the transition probabilities of $\{ L_{k} \}_{k=1}^{\infty}$ are given by

\begin{equation*}\label{pr2r}
r_{ij} = \left \{ \begin{array}{ll} 1-p_{01}^{(i+1)}(1- \alpha) & j=1 \\ p_{01}^{(i+1)}(1 - \alpha)^{j-1}p_{11}^{j-2}(1-p_{11}(1-\alpha)) &  j \geq 2 \end{array} \right.,
\end{equation*}

where $P_{01}^{(j)}$ is the $j$-step transition probability which is equal to $\omega_{0} - \omega_{0}(p_{11}-p_{01})^{j}$. Let $\mathbf{R} = \{ r_{ij} \}$ denote the transition matrix of $\{ L_{k} \}_{k=1}^{\infty}$ and let $\mathbf{R}(:,k)$ denote the k-th column of $\mathbf{R}$. We have 
\begin{equation}\label{pr2R1}
\mathbf{1} - \mathbf{R}(:,1) = \frac{\mathbf{R}(:,2)}{1-p_{11}(1-\alpha)},
\end{equation}
and
\begin{equation}\label{pr2R2}
\mathbf{R}(:,k) = \mathbf{R}(:,2)(p_{11}(1-\alpha))^{k-2} \mbox{ for }  k \geq 2,
\end{equation} 
where $\mathbf{1}$ is the unit column vector $[1,1,\cdots]^{T}$. We denote $\Lambda = [\lambda_{1}, \lambda_{2}, \cdots]$ as the stationary distribution of $\{ L_{k} \}_{k=1}^{\infty}$, i.e. $\Lambda \mathbf{R} = \Lambda$. Thus we have: 
\begin{equation}\label{pr2lam}
[\lambda_{1}, \lambda_{2}, \cdots] \mathbf{R}(:,k) = \lambda_{k}.
\end{equation}

Combining \eqref{pr2R1}, \eqref{pr2R2} and \eqref{pr2lam} results in
\begin{equation}\label{pr2lam2}
\lambda_{1}  = 1 - \frac{\lambda_{2}}{1-p_{11(1-\alpha)}}, \mbox{   }  \lambda_{k} = \lambda_{2}(p_{11}(1-\alpha))^{k-2}.
\end{equation}
Substituting \eqref{pr2lam2} into \eqref{pr2lam} and solving for $\lambda_{2}$, we get $\lambda_{2} = \overline{\omega}(1 - p_{11}(1-\alpha))$, where $\overline{\omega}$ is given in \eqref{T1O}. From \eqref{pr2lam2}, we can find the stationary distribution as
\begin{equation}\label{pr2dist}
\lambda_{k} = \left \{ \begin{array}{ll} 1-\overline{\omega}, & k=1 \\ \overline{\omega}(1-p_{11}(1-\alpha))(p_{11}(1-\alpha))^{k-2} & k>1 \end{array}  \right..
\end{equation}
Using \eqref{pr2dist} to compute $L^{m}(\alpha) = \sum_{k=1}^{\infty}k\lambda_{k}$ results in \eqref{T1L}.
\end{proof}

Using the results of Theorem~1, we can show that $U^{m}(\alpha)$ is a decreasing function of $\alpha$, and thus an attacker can lower the target radio's throughput (which is employing myopic sensing) by increasing the attack probability $\alpha$. However, as we discussed in Section~2.2, increasing $\alpha$ increases the probability that the attack is detected.

\subsection{Analysis of the Softmax Policy}\label{ssec:2softmax}

The \emph{softmax} action selection policies are randomized policies where, at time $t$, the action $a_{t}$ is chosen at random by the user according to some probability distribution giving more weight to actions which have performed well in the past. The greedy action is given the highest selection probability, but all the others are ranked and weighted according to their accumulated rewards \cite{bahrak:en:sb:98}. The most common softmax action selection method uses a Gibbs or Boltzman distribution for the action selection probabilities. It chooses action $a$ at time slot $t$, with probability
\begin{equation*}
p_{a}(t) = \frac{e^{\omega_{a}(t) / \tau}}{\sum_{i=1}^{N} e^{\omega_{i}(t) / \tau}},
\end{equation*}
where $\tau$ is a positive parameter called the \emph{temperature} and controls the greediness of the policy. High temperatures cause all the actions to be all equiprobable while low temperatures cause a high probability for greedy action, and in the limit as $\tau \rightarrow 0$, the softmax policy becomes equivalent to the myopic policy.

 In this section, for simplicity and ease of computation, we use a Bernoulli distribution instead of a Boltzmann distribution, i.e., we choose action $a$ at time slot $t$ with probability
 \begin{equation}\label{main}
p_{a}(t) = \left \{ \begin{array}{lcl} q &  \mbox{if} & a = \arg \max_{i=1,2} \omega_{i}(t)  \\ 1-q & \mbox{if} & a= \arg \min_{i=1,2} \omega_{i}(t)  \end{array} \right..
\end{equation}

We define the \emph{main probability} $q$ as the probability of taking greedy action (i.e. selecting the channel that has the highest $\omega_{i}$). According to the definition of $q$ and the softmax policy, we have $ 0.5 \leq q \leq 1$ when $N=2$. Also note that for $q = 1$, the softmax policy reduces to myopic policy, i.e., myopic policy is a special case of softmax policy.

The attacker's optimal strategy for attacking a channel selection system employing the myopic policy is simple: only attack the channel that has the biggest belief value, since the user does not transmit on other channels. The attackerÕs optimal strategy against the softmax policy is not so straightforward.

\begin{figure}[h]
 \centering
 \subfigure[attacker's strategy with division probability d = 1]{
  \includegraphics[scale=0.6]{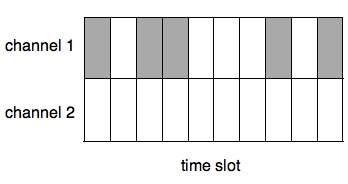}
   \label{fig:subfig1}
   }
 \subfigure[attacker's strategy with division probability d = 0.6]{
  \includegraphics[scale=0.6]{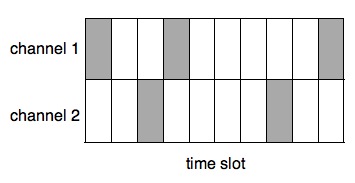}
   \label{fig:subfig2}
   }
 \label{fig:subfigureExample}
 \caption[]{%
  attack strategy examples for fixed attack probability $\alpha=0.5$}
\end{figure}

As mentioned in Section~2.2, the $\alpha$-optimal strategy for an attacker is a strategy that minimizes the throughput of a cognitive radio while keeping the attack probability $\alpha$ fixed. Knowing that the softmax policy uses a fixed main probability $q$ for channel selection (i.e., it uses \eqref{main} for channel selection), the attacker divides its attacks between the two channels. We define $d$ as the conditional probability of channel 1 (the greedy option for the policy) being attacked at a given timeslot, assuming that the timeslot is attacked and call it the \emph{division probability}. Figure~2 illustrates the concept of the division probability (the slots colored in gray are the time slots in which an attacker jams a channel). The attacker chooses $d$ such that a cognitive radioÕs throughput is minimized. On the other hand the channel selection system exploits its knowledge about the optimal strategy of the attacker to select the main probability $q$ such that the throughput is maximized. Assume that $U^{s}(q,d)$ define the throughput of the radio when the softmax policy uses a main probability of $q$ and the attacker exploits the division probability $d$, constructing optimal attack strategy and its corresponding optimal channel selection probability distribution can be formulated as the following optimization problems:

{\bf Optimization Problem 1:}
\begin{align*}
&d^{*} = \min_{d} U^{s}(q,d)\\
&\mbox{s.t.          } 0 \leq d \leq 1
\end{align*}

{\bf Optimization Problem 2:}
\begin{align*}
&q^{*} = \max_{q} U^{s}(q,d^{*})\\
&\mbox{s.t.          } 0.5 \leq q \leq 1
\end{align*}

The steady-state throughput of the softmax policy is given by:
\begin{equation*}
U^{s} = \lim_{T \rightarrow \infty} \frac{V^{s}_{1:T}(\Omega(1))}{T},
\end{equation*}
where $V^{s}_{1:T}(\Omega(1))$ is the expected total reward obtained in $T$ slots under the softmax policy when the initial belief vector is $\Omega(1)$. As shown in Section~3.1, we only need the average TP length to compute the throughput for the softmax policy.
Suppose that the attacker uses its optimal strategy. An analysis similar to what we did in section~2, results in the following theorem.

 \newtheorem{T3}[theorem]{Theorem}
 \begin{T3}
For $N=2$, the average TP average length for a softmax policy with main probability $q$ is given by:
\begin{equation*}
L^{s}(q,d) = q L^{m}(\alpha d) + (1-q)L^{n}(\alpha(1-d)),
\end{equation*}
where $L^{m}(\cdot)$ is the function defined in \eqref{T1L} and $L^{n}(x) = 1 + \frac{p_{01}(1-x)}{1-p_{11}(1-x)}$.
\end{T3}
 \begin{proof}
Using a procedure similar to the one used in the proof of Theorem~1, we  can readily show that if the channel selection algorithm always selects the channel with the smaller belief value (which is the opposite of a greedy action), then the stationary distribution of $\{ L_{k} \}_{k=1}^{\infty}$ is 
\begin{equation}\label{pr3lam}
\lambda_{k} = \left \{ \begin{array}{ll} 1-p_{01}(1-\alpha), & k=1 \\ p_{01}(1-\alpha)(1-p_{11}(1-\alpha))(p_{11}(1-\alpha))^{k-2} & k>1 \end{array}  \right.
\end{equation}
Using \eqref{pr3lam} to compute $L^{n}(\alpha) = \sum_{k=1}^{\infty}k\lambda_{k}$ results in 
\begin{equation*}
L^{n}(\alpha) = 1 + \frac{p_{01}(1-\alpha)}{1-p_{11}(1-\alpha)}.
\end{equation*}
By using the BayesÕ rule and the statement of Theorem~1, we can obtain the statement of Theorem~2.
 \end{proof}

To quantify how much randomness is added to the channel selection system by the softmax policy, we use the entropy of the channel selection probability distribution, $\mathcal{H}$. For the $N=2$ case, because we use Bernoulli distribution, $\mathcal{H} = -(q \ln(q) + (1-q)\ln(1-q))$. By changing $q$ from $1$ to $0.5$, the entropy increases from $0$ to its maximum value $\ln(2)$. 

We denote $U^{\pi}(\alpha)$ as the throughput of the cognitive radio, when it uses policy $\pi$ for channel selection and the attacker uses its $\alpha$-optimal strategy. We use the following definitions to quantify the robustness and the performance of policy $\pi$:

 \newtheorem{D1}[definition]{Definition}
\begin{D1}
The {\bf robustness} of a policy $\pi$ for a channel selection system under an $\alpha$-optimal attack is: $R^{\pi}(\alpha) = 1- \frac{U^{\pi}(0) - U^{\pi}(\alpha)}{U^{\pi}(0)}$.
\end{D1}

 \newtheorem{D2}[definition]{Definition}
\begin{D2}
The {\bf performance} of a policy $\pi$ for a channel selection system under an $\alpha$-optimal attack is: $P^{\pi} = U^{\pi}(0)$.
\end{D2}

\subsection{Numerical Results for two Channels}\label{ssec:2num}
Using the results of Theorem 2, it can easily be shown that for $\alpha = 0$, $q^{*} = 1$, i.e., a non-adversarial environment the optimal softmax policy is equivalent to the myopic policy. Solving the optimization problems 1 and 2 for other values of $\alpha$ is straightforward due to the problem's small solution space. Figure~3 shows the solutions to this problem for a range of attack probabilities. Because myopic sensing is a special case of softmax sensing, it is obvious that  $U^{s} \geq U^{m}$. Numerical results illustrated in Figure~3 shows that except for small values of $\alpha$ ($\alpha < 0.1$),  $U^{s}$ is strictly greater than $U^{m}$, i.e., softmax sensing outperforms myopic sensing.  
  \begin{figure}[t]
\centering 
    \includegraphics[width=4in]{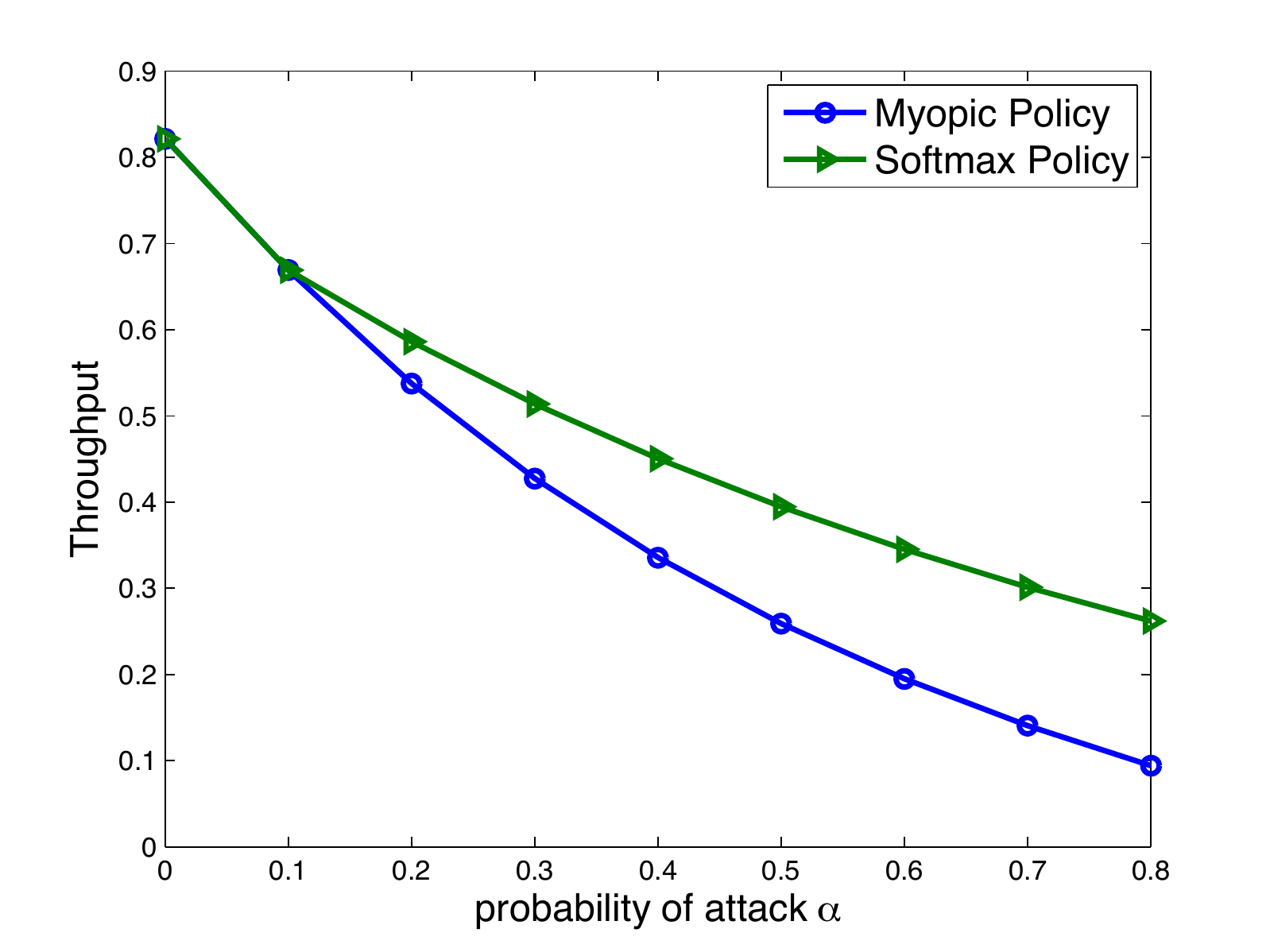}
\caption{Throughput vs. attack probability for $N=2$.}
\label{qopt1}
\end{figure}

Figure~4 shows the trade-off between the robustness and the performance of the system for fixed attack probability $\alpha = 0.5$. As it can be seen, increasing the randomness that is used in the system by the softmax policy, increases the policy's robustness but it would decrease the policy's performance when no attacker exists.
  \begin{figure}[t]
\centering 
    \includegraphics[width=7in]{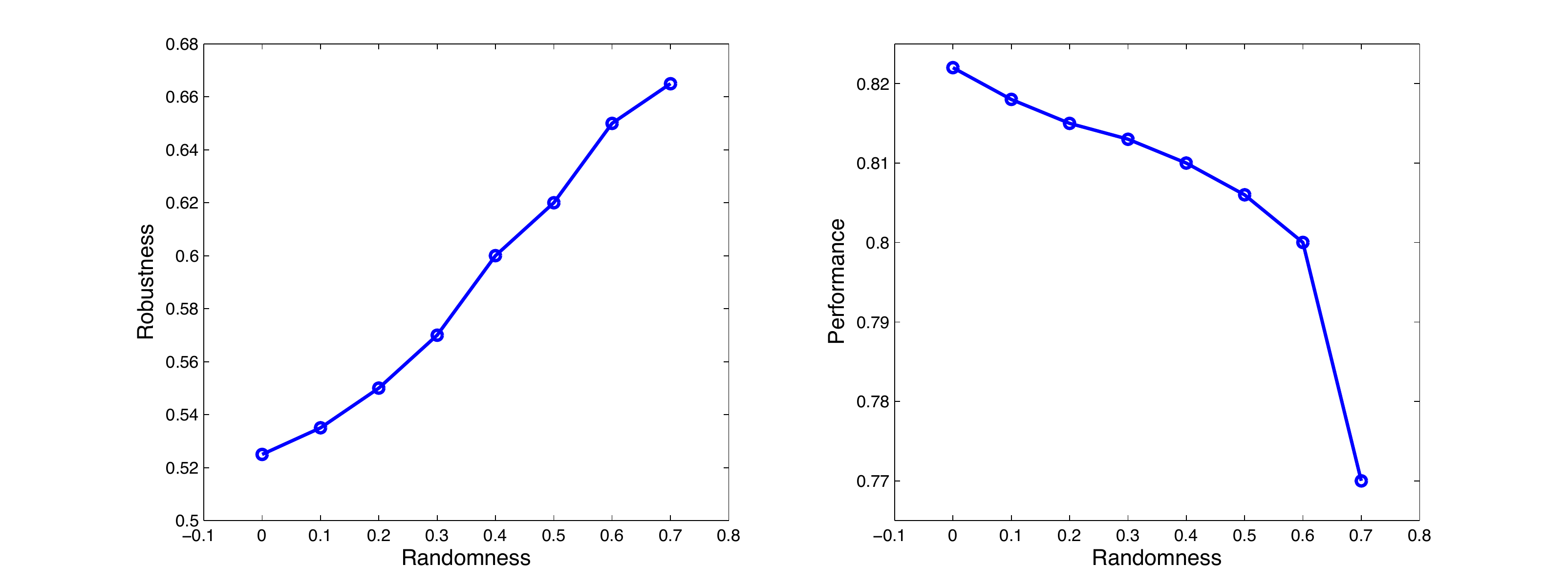}
\caption{Performance and robustness vs. randomness for $N=2$.}
\label{qopt2}
\end{figure}

\section{Sensing Policies with More than Two Channels in an Adversarial Environment}\label{sec:more}

As mentioned in \cite{bahrak:en:zkl:08}, because $L_{k}$ is a random process with higher-order memory, obtaining a closed-form expression of throughput for $N>2$ channels with different statistics is very difficult. Nevertheless we can show that a softmax policy that uses a Boltzmann distribution with an intelligent choice for the temperature, $\tau$, can outperform myopic policy for all possible strategies of the attacker, including its optimal strategy. 

Assume that the channel selection system selects one of the $N$ channels at each time slot. When a channel is selected, the system keeps using that channel until a primary user begins to transmit on the channel. Suppose $\Omega(t) = (\omega_{1}(t), \omega_{2}(t) , \cdots , \omega_{N}(t))$ is the belief vector of the system at time $t$ and $Q(t) = (q_{1}(t), q_{2}(t) , \cdots , q_{N}(t))$ is the corresponding probability vector, i.e. $q_{i}(t) = f_{i}(\Omega(t))$ is the probability of selecting channel $i$ at time $t$ where $f(\cdot)$ is a function that maps the belief vector into a probability value. For the sake of notation simplicity, we omit $t$ in the belief and and probability vectors henceforth. Without loss of generality, assume that $\omega_{N} \geq \omega_{N-1} \geq \cdots \geq \omega_{1}$ and consequently $q_{N} \geq q_{N-1} \geq \cdots \geq q_{1}$. As we mentioned in Section~2, the $\alpha$-optimal strategy for the attacker is a strategy that minimizes the throughput while keeping the attack probability $\alpha$ fixed. In order to do so, the attacker needs to attack channel $i$ with probability $\alpha d_{i}$, where $d_{i}$ is the division probability for channel $i$ (i.e., the conditional probability of channel $i$ being attacked at a given timeslot, assuming that the timeslot is attacked) which is a function of $Q$ and $\Omega$. Also to keep $\alpha$ fixed, we need to have $\sum_{i=1}^{N} d_{i} = 1$. The following theorem states the optimization problem that the attacker needs to solve to find its optimal strategy.

 \newtheorem{T6}[theorem]{Theorem}
 \begin{T6}
In order to find its $\alpha$-optimal strategy, the attacker needs to solve the following equality-constrained convex optimization problem:\\
{\bf Optimization Problem 3.}
\begin{equation*}
\min_{d_{1} , \cdots , d_{N}}  \sum_{i=1}^{N}  q_{i} L(\omega_{i}, \alpha d_{i})
\end{equation*}
\begin{equation*}
\mbox{s.t.} \sum_{i=1}^{N} d_{i} = 1,
\forall i : d_{i} \geq 0
\end{equation*}
where 
\begin{equation}\label{T6L}
L(\omega_{i},\alpha) = 1 + \frac{\omega_{i}(1-\alpha)}{1-p^{i}_{11}(1-\alpha)}.
\end{equation}
\end{T6}
 \begin{proof}
In order to minimize the throughput, attacker needs to minimize the expected value for $L_{k}(\omega)$. We know that the attacker attacks this channel with probability $\alpha d_{i}$, so we have:
\begin{equation*}
Pr\{L_{k}(\omega_{i}) = l\} = \left \{ \begin{array}{ll} 1-\omega_{i}(1-\alpha d_{i}), & l=1 \\ \omega_{i}(1-\alpha d_{i})(p^{i}_{11}(1-\alpha d_{i}))^{l-2}p^{i}_{10}, & l>1 \end{array} \right.
\end{equation*}
It can easily be shown that
\begin{equation*}
 E[L_{k}(\omega)|\mbox{channel } i] = 1 + \frac{\omega_{i}(1-\alpha d_{i})}{1-p^{i}_{11}(1-\alpha d_{i})}.
\end{equation*}
So we have:
\begin{align*}
E[L_{k}(\omega)] &= E[E[L_{k}(\omega)|\mbox{channel } i]] \\
&= \sum_{i=1}^{N} q_{i}(1 + \frac{\omega_{i}(1-\alpha d_{i})}{1-p^{i}_{11}(1-\alpha d_{i})}).
\end{align*}
To prove that the $E[L_{k}(\omega)] $ is a convex function of $d_{i}$'s, we compute the Hessian matrix of this function: $H_{N\times N} = [h_{ij}]$. We can see that the Hessian of this function is a diagonal matrix where:
\begin{equation*}
h_{ii} = \frac{2q_{i}\omega_{i}\alpha^{2}p^{i}_{11}}{(1-p^{i}_{11}(1-\alpha d_{i}))^{3}}.
\end{equation*}
It is obvious that we have $\forall i : h_{ii} \geq 0$, thus the matrix $H$ is positive semidefinite and as a result  $E[L_{k}(\omega)]$ is a convex function.
 \end{proof}

The above equality-constrained convex optimization problem can be solved by using elimination and Newton's method in $\frac{(N+1)^{3}}{3}$ steps \cite{bahrak:en:bv:04}. However, solving this optimization problem requires attacker's exact knowledge about the channel selection strategy of the system and the statistics of all channels during a long period of time. Obviously, acquiring such knowledge is not feasible under most circumstances. Therefore instead of using the $\alpha$-optimal strategy, the attacker can use alternative, simpler strategies. Each of these strategies differ in terms of the amount of knowledge on the channel selection system that is required. These strategies are:

\begin{itemize}
\item Greedy Strategy: The attacker only knows the best channel for transmission at each time slot and attacks this channel, i.e., $d_{N} = 1$ and $d_{i} = 0$ for $1 \leq i \leq N-1$. This strategy is the optimal attack strategy when the channel selection system uses a myopic policy.
\item Uniform Strategy: The adversary attacks all channels equiprobably, i.e., $d_{i} = \frac{1}{N}$ : $1 \leq i \leq N$. This strategy can be used when the attacker does not have any knowledge about the channel selection system.
\item $\Omega$ Strategy: The attacker only knows the channel statistics $\Omega = (\omega_{1} , \cdots , \omega_{N})$ and has no knowledge about the channel selection policy of the system. It has a Boltzmann distribution with an arbitrary temperature $\tau_{a}$, i.e., $d_{i} = \frac{e^{\omega_{i} / \tau_{a}}}{\sum_{j=1}^{N} e^{\omega_{j} / \tau_{a}}}$ : $1 \leq i \leq N$. Simulation results show that when the attack probability, $\alpha$, is large, this strategy inflicts approximately the same effect as the $\alpha$-optimal strategy.
\end{itemize}

In order to find the best channel selection strategy, the system assumes the worst-case attack scenario when the attacker uses its $\alpha$-optimal strategy, which is the solution to the Optimization Problem~3: $d_{i}^{*} = f_{i}^{*}(Q,\Omega)$. The channel selection system needs to solve the following optimization problem in order to maximize the cognitive radio's throughput.

{\bf Optimization Problem 4.}
\begin{equation*}
\max_{q_{1} , \cdots , q_{N}}  \sum_{i=1}^{N}  q_{i} L(\omega_{i}, \alpha f_{i}^{*}(Q,\Omega))
\end{equation*}
\begin{equation*}
\mbox{s.t.} \sum_{i=1}^{N} q_{i} = 1,
\forall i : q_{i} \geq 0
\end{equation*}

Finding the global optimal solution of the non-linear optimization problem 4, gives us the amount of randomness that we need to add to the system in order to minimize the attack's effect and maximize the throughput. 

We can also show that \emph{for all possible attack strategies, including the $\alpha$-optimal strategy, a softmax policy that uses a Boltzmann distribution with a well-chosen temperature outperforms myopic policy}.

 \newtheorem{T4}[theorem]{Theorem}
 \begin{T4}
When the channel selection system is consisted of more than two identical channels, for all attack strategies that the adversary may employ, with a fixed attack probability $\alpha > \frac{(\omega_{0} - p_{01})N}{(\omega_{0}N - p_{01})}$, a softmax policy that uses a Boltzmann distribution with temperature
\begin{equation}\label{tau}
 \tau > \frac{\omega_{0} - p_{01}}{\ln \frac{p_{01}(N-\alpha)}{\omega_{0}N(1-\alpha)}},
 \end{equation} 
 achieves a greater throughput than a myopic policy.
\end{T4}
 \begin{proof}
Let $\omega_{N} \geq \omega_{N-1} \geq \cdots \geq \omega_{1}$ denote the belief values of all channels in the first slot of the k-th TP. For the myopic policy, the length $L_{k}$ of this TP has the following distribution.
\begin{equation*}
Pr\{L_{k}(\omega_{N}) = l\} = \left \{ \begin{array}{ll} 1-\omega_{N}(1-\alpha), & l=1 \\ \omega_{N}(1-\alpha)(p_{11}(1-\alpha))^{l-2}p_{10}, & l>1 \end{array} \right.
\end{equation*}
And for the softmax policy, the length $L_{k}$ of this TP has the following distribution.
\begin{equation*}
Pr\{L_{k}(\overline \omega) = l\} = \left \{ \begin{array}{ll} 1-\overline \omega, & l=1 \\ \overline \omega(p_{11}(1-\alpha \overline d))^{l-2}p_{10}, & l>1 \end{array} \right.,
\end{equation*}
where $\overline \omega = \sum_{i=1}^{N} q_{i}\omega_{i}(1-\alpha d_{i})$ and $\overline d = \sum_{i=1}^{N}q_{i}d_{i}$ in which $q_{i} = \frac{e^{\omega_{i} / \tau}}{\sum_{j=1}^{N} e^{\omega_{j} / \tau}} = \frac{1}{\sum_{j=1}^{N} e^{(\omega_{j} - \omega_{i}) / \tau}}$ are the action selection probabilities that follow a Boltzmann distribution.
It is readily observable that if $\overline \omega \geq \omega_{N}(1-\alpha)$, then $L_{k}(\overline \omega)$ stochastically dominates $L_{k}(\omega_{N})$ and consequently the throughput of the softmax policy would be greater than the throughput of the myopic policy. We use the fact that $\forall i$, 
$p_{01} \leq \omega_{i} \leq \omega_{0}$ which results in 
\begin{equation}\label{taucond}
\frac{1}{N e^{(\omega_{0} - p_{01})/\tau}} \leq q_{i} \leq \frac{1}{N e^{(p_{01} - \omega_{0})/\tau}},
\end{equation}
and also the fact that $q_{N} > \frac{1}{N}$ and $d_{N} > \frac{1}{N}$.

Now we show that  $\overline \omega \geq \omega_{N}(1-\alpha)$. Using \eqref{taucond} and \eqref{tau}, we have:
\begin{align*}
\overline \omega &= \omega_{N} q_{N} (1-\alpha d_{N}) + \sum_{i=1}^{N-1} q_{i} \omega_{i} (1-\alpha d_{i}) \\
& \geq \omega_{N} q_{N} (1-\alpha d_{N}) + \sum_{i=1}^{N-1} p_{01} q_{i} (1-\alpha d_{i})\\
& \geq \omega_{N} q_{N} (1-\alpha d_{N}) + \sum_{i=1}^{N-1} p_{01} \frac{1}{N e^{(\omega_{0} - p_{01})/ \tau}} (1-\alpha d_{i})\\
& \geq \omega_{N} q_{N} (1-\alpha d_{N}) + \frac{p_{01}}{N} \frac{\omega_{0} N (1-\alpha)}{p_{01}(N-\alpha)}(N-1-\alpha(1- d_{N})) \\
& \geq \omega_{N} (q_{N} (1-\alpha d_{N}) +  \frac{(1-\alpha)}{(N-\alpha)}(N-1-\alpha(1- d_{N})) )\\ 
& \geq \omega_{N} (\frac{1}{N} (1-\alpha) +  \frac{(1-\alpha)}{(N-\alpha)}(N-1-\alpha(1- \frac{1}{N})) )\\ 
& = \omega_{N}(1-\alpha)
\end{align*}.

\end{proof}

\subsection{Numerical Results for More than Two Channels}

  \begin{figure}[t]
\centering 
    \includegraphics[width=4in]{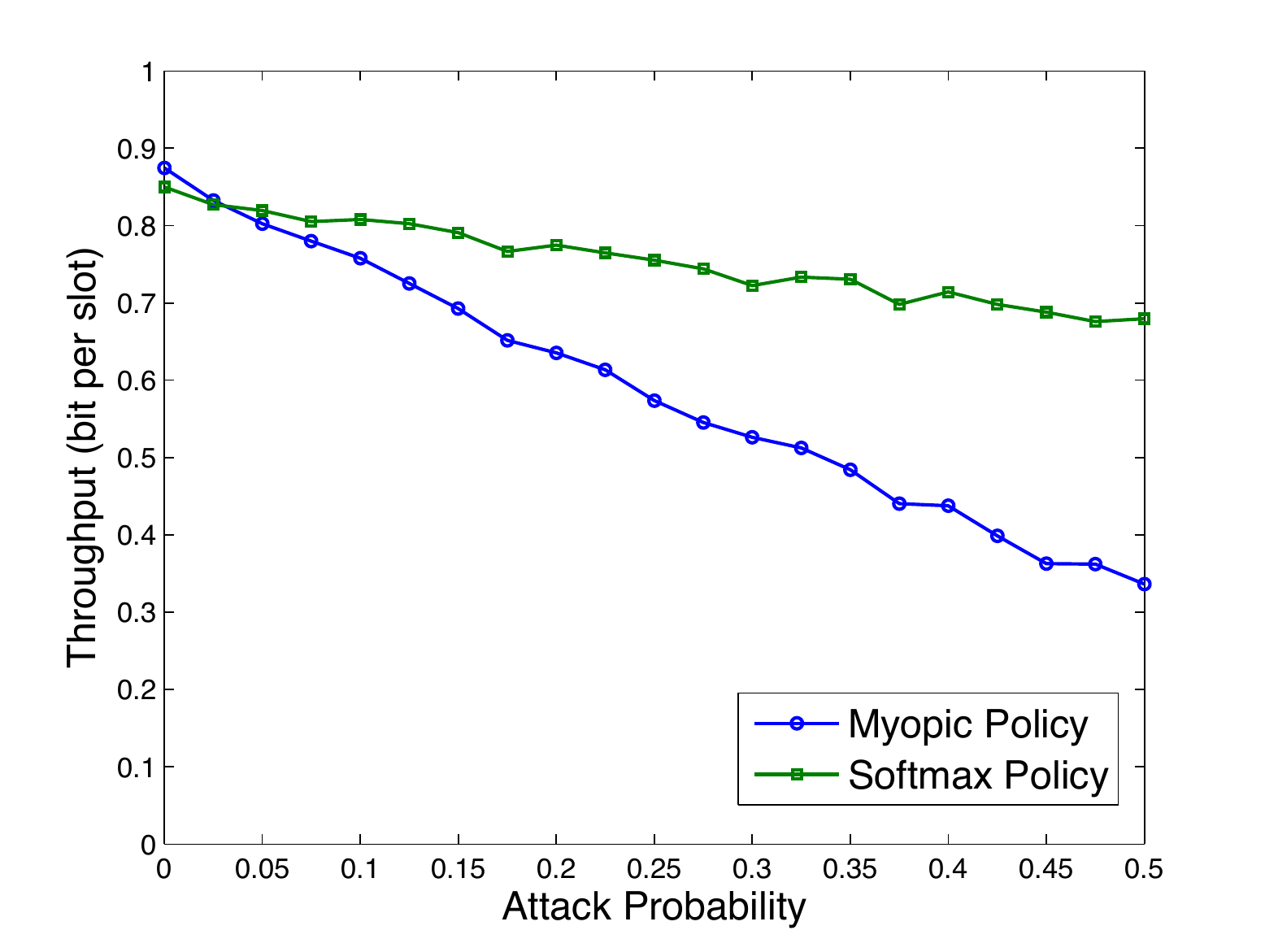}
\caption{Throughput vs. attack probability for N=4.}
\label{sim1}
\end{figure}

  \begin{figure}[t]
\centering 
    \includegraphics[width=4in]{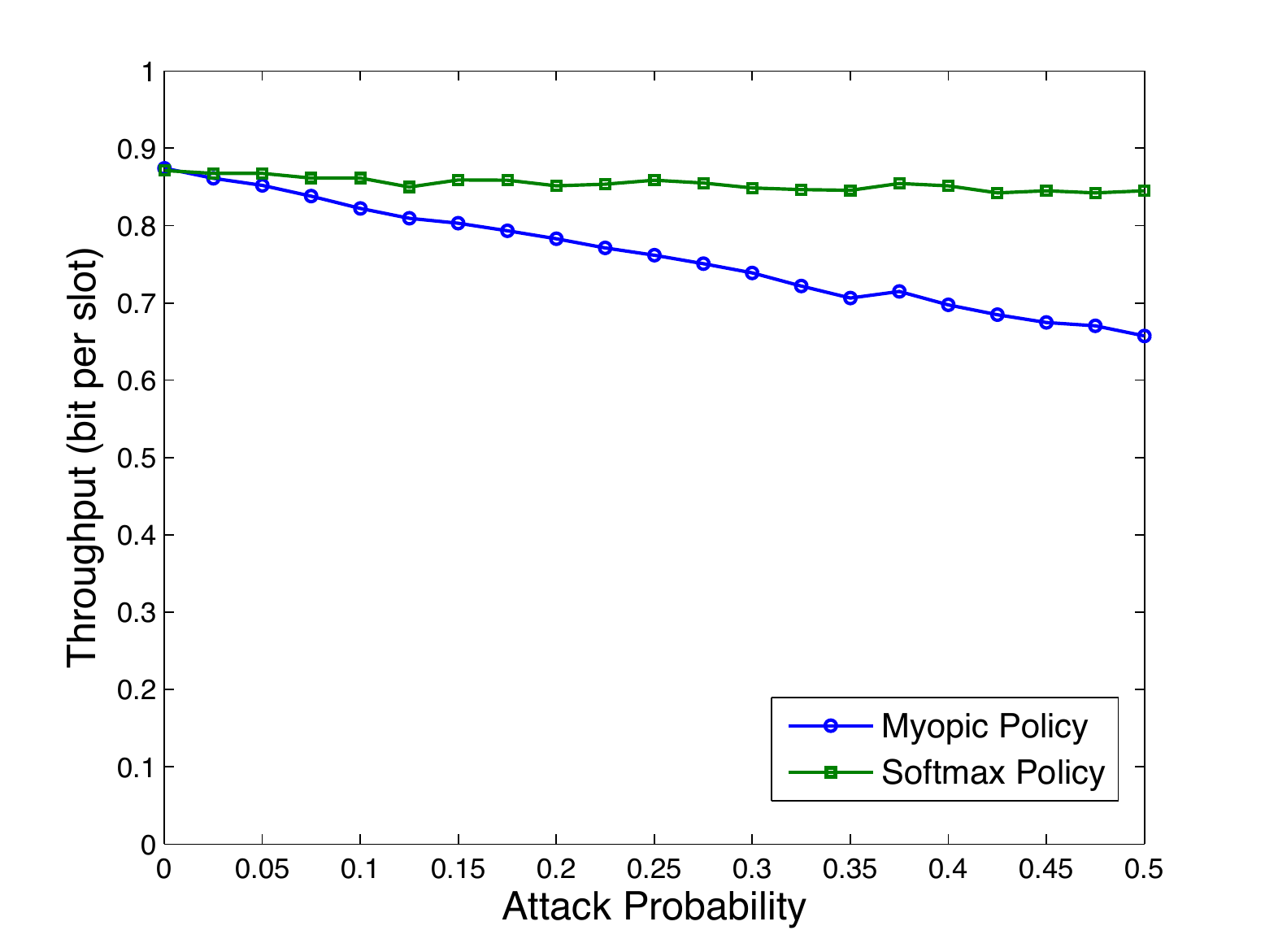}
\caption{Throughput vs. attack probability for N=10.}
\label{sim2}
\end{figure}

We simulated the channel selection system to evaluate the performance of myopic sensing and softmax sensing for more than two channels. Figures~5 and 6 show the throughput of the channel selection system versus the attack probability for myopic and softmax policies when 4 channels and 10 channels are available, respectively. To observe the effect of the number of channels on our scheme, we used identical channels with transition probabilities $p_{11} = 0.9$, $p_{10} = 0.1$, $p_{00} = 0.8$ and $p_{01} = 0.2$ to obtain the results in Figures~5 and 6. Softmax policies in these figures use a Boltzmann distribution with fixed temperature $\tau = 2$ for channel selection. 

As it can be seen, the throughput drop caused by the increase in the attack probability is more severe for the myopic policy. Comparing the slope of lines in Figures~5 and 6, we can see that increasing the number of channels from 4 to 10 makes the softmax policy more robust against the belief manipulation attacks. In Figure~6, the softmax policy's drop in throughput is barely noticeable even as the attack probability is increased to $0.5$.

Figures 5 and 6 also show that in a non-adversarial environment (i.e. when $\alpha = 0$), the performance of myopic policy is better than the performance of a softmax policy.  These figures confirm the findings of \cite{bahrak:en:zkl:08} in which the authors showed that the optimal policy in non-adversarial environments is the myopic policy.  

\begin{table}
\centering
  \caption{Transition Probabilities}
\begin{tabular}{  c | c | c | c | c }
\hline			
     & $p_{11}$ & $p_{10}$ & $p_{00}$ & $p_{01}$\\
     \hline
  Channel 1 & 0.9 & 0.1 & 0.8 & 0.2\\
  Channel 2 & 0.95 & 0.05 & 0.8 & 0.2\\
  Channel 3 & 0.9 & 0.1 & 0.85 & 0.15\\
  Channel 4 & 0.95 & 0.05 & 0.85 & 0.15\\
\hline  
\end{tabular}
\end{table}

To compare different attack strategies and to investigate the amount of required randomness in the channel selection system, we used the more realistic model of non-identical channels with different transition probabilities. Figure 7 shows the effectiveness of the different attack strategies for different attack probabilities in a system of 4 non-identical channels with transition probabilities shown in Table~1. 

  \begin{figure}[t]
  \centering
    \includegraphics[width=4in]{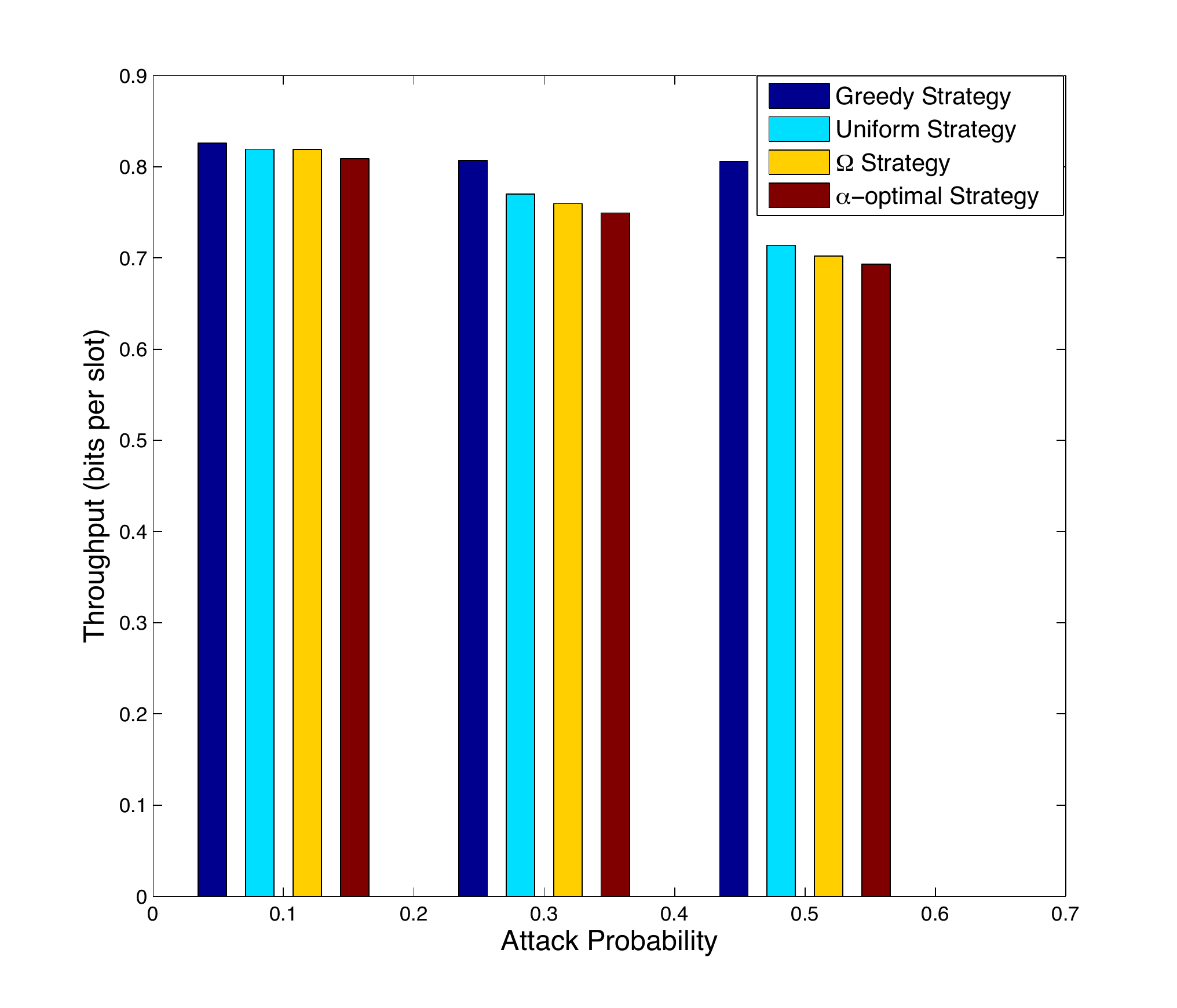}
\caption{Comparing different attack strategies.}
\label{sim3}
\end{figure}

As can be seen in Figure~7, the $\Omega$ strategyÕs performance is close to that of the $\alpha$-optimal strategy for large values of $\alpha$. We can also see from the figure that the greedy strategy is the worst strategy among these four different strategies.

  \begin{figure}[t]
  \centering
    \includegraphics[width=4in]{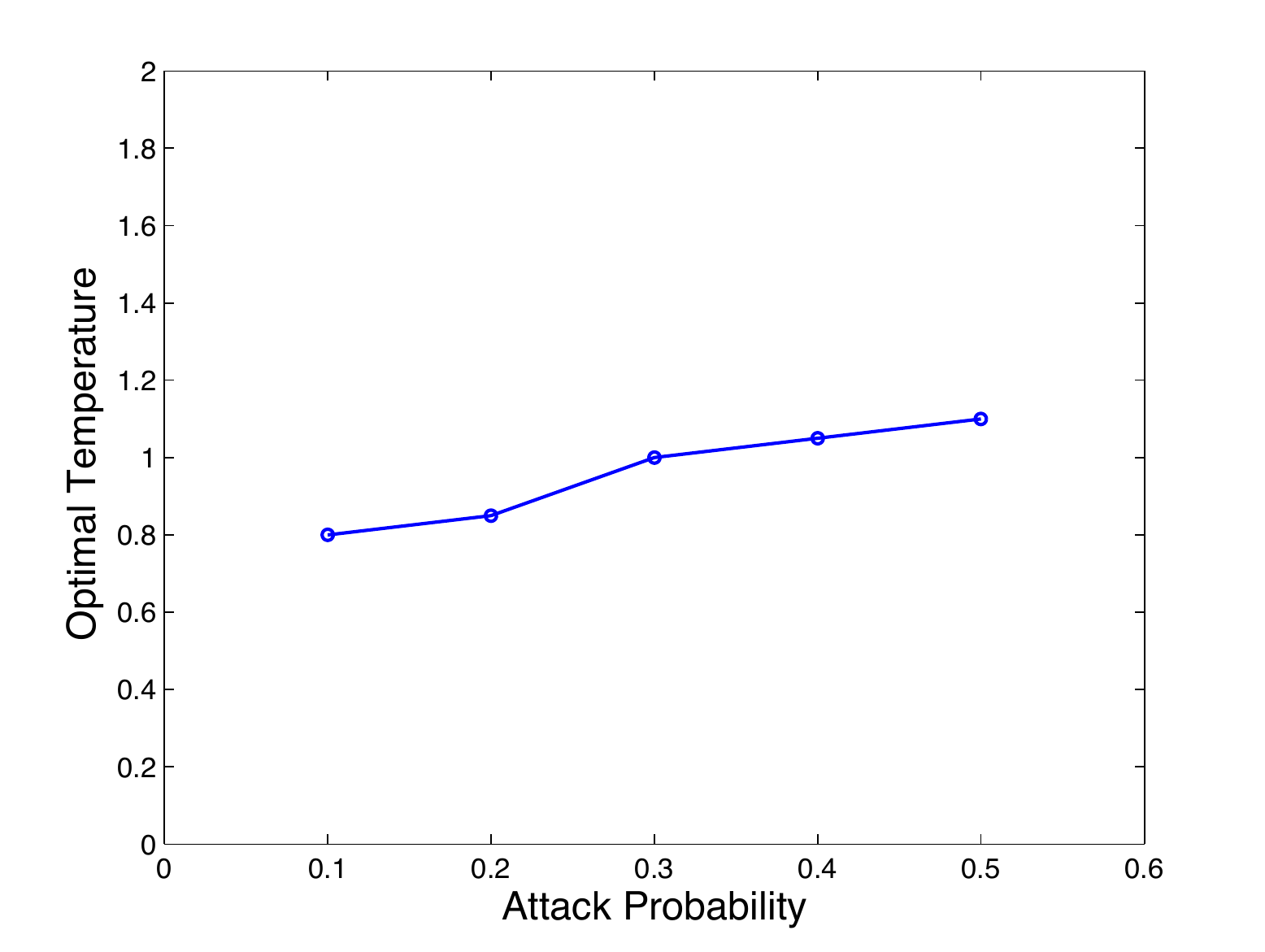}
\caption{Optimal amount of randomness vs. attack probability}
\label{sim3}
\end{figure}
It is well known that the temperature of a Boltzmann distribution, $\tau$, is a measure of the amount of randomness in the distribution, which can be inferred from the fact that the entropy of the Boltzmann distribution is an increasing function of $\tau$ \cite{bahrak:en:om:93}. Figure~8 plots the optimal temperature value for the channel selection system when the attacker uses the $\alpha$-optimal strategy. These temperature values were obtained by solving Optimization Problem~4. From the figure, we can observe that the channel selection system needs to increase the randomness (i.e., $\tau$) in its learning process as the attack probability (i.e., $\alpha$) is increased to minimize the effect of the attack.

\begin{figure}[t]
  \centering
    \includegraphics[width=4in]{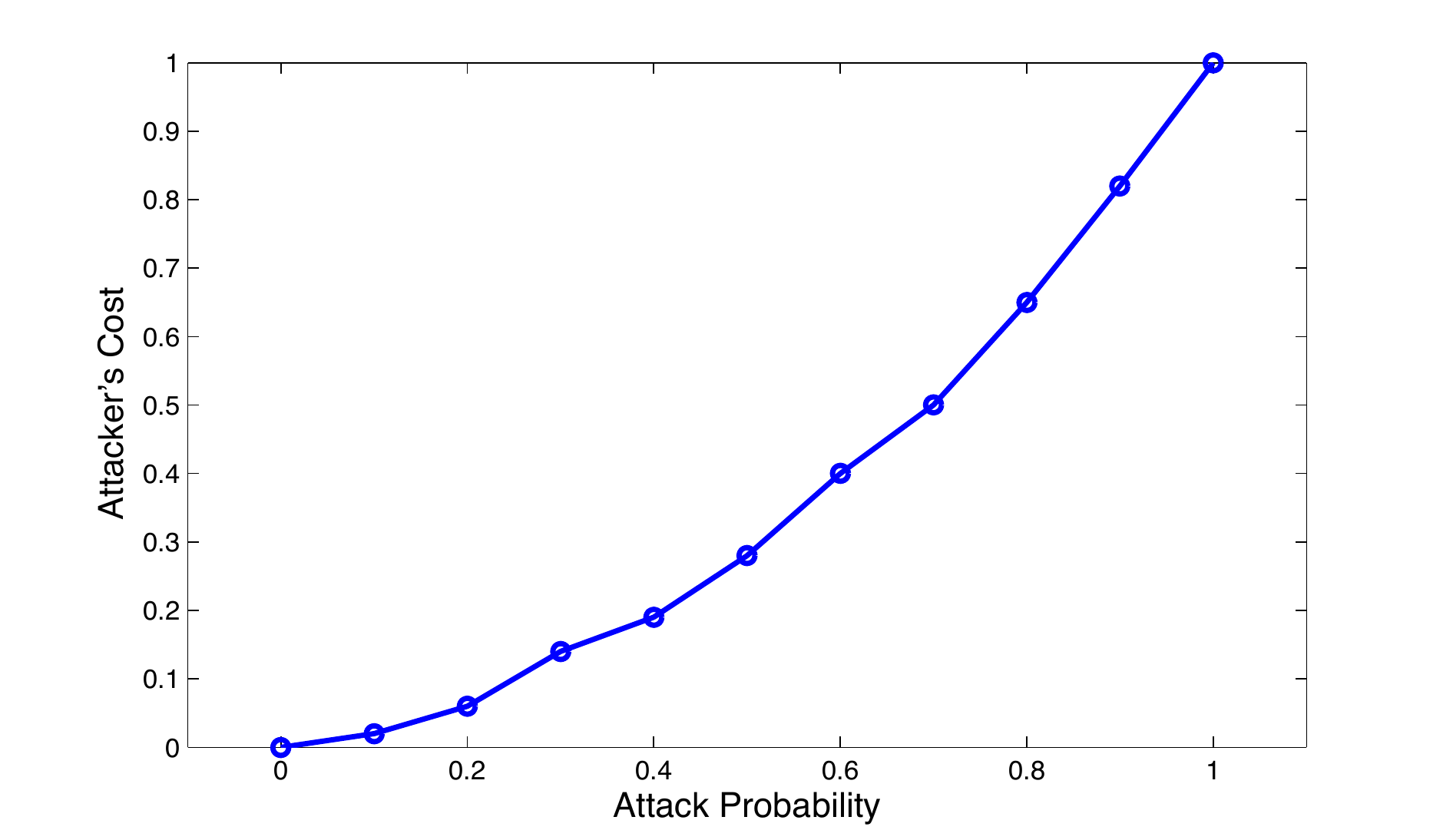}
\caption{Attacker's cost vs. attack probability}
\label{fig:cost}
\end{figure}

We observed the effect of increasing attack probability on performance of the channel selection system. Figure~\ref{fig:cost} plots the attacker's cost using the equation~\ref{attackcost} versus the attack probability. This figure shows that though increasing the attack probability reduces the performance of the channel selection system significantly, it also increases the  attacker's cost with a high rate.

\section{Conclusion}\label{sec:con}
In this paper, we analyzed the security of a reinforcement learning algorithm that is used for solving the channel selection problem. We proposed a sensing policy that uses some level of randomness in the decision function to hide information about the learning algorithm. The obtained theoretical and simulation results show that the proposed mitigation technique can cause a dramatic improvement in the robustness of the channel selection process to adaptive jamming attacks. This countermeasure is applicable to other cognitive radio applications that use the same type of machine learning algorithm.


\balance


\begin{thebibliography}{?}
        
\bibitem{bahrak:en:czs:08}
	Y. Chen, Q. Zhao, and A. Swami, ``Joint design and separation principle for opportunistic spectrum access in the presence of sensing errors," IEEE Transactions on Information Theory, vol. 54, no. 5, May, 2008.
	
\bibitem{bahrak:en:zkl:08}
	 Q. Zhao, B. Krishnamachari, and K. Liu, ``On myopic sensing for multi- channel opportunistic access: structure, optimality, and performance," IEEE Transactions on Wireless Communications, 2008.
	 
\bibitem{bahrak:en:alj:09}
	  S.H. Ahmad, M. Liu, T. Javidi, Q. Zhao and B. Krishnamachari, ``Optimality of Myopic Sensing in Multi-Channel Opportunistic Access" , in IEEE Transactions on Information Theory, vol. 55, No. 9, pp. 4040-4050, September, 2009.
	
	   
\bibitem{bahrak:en:al:09} 
	   S.H.A.Ahmad and M.Liu,``Multi-channel opportunistic access: a case of restless bandits with multiple plays," Allerton Conference, 2009.
	   
 \bibitem{bahrak:en:lz:08}
	   K. Liu and Q. Zhao, ``A restless bandit formulation of opportunistic access: indexablity and index policy," Proc. of the 5th IEEE Conference on Sensor, Mesh and Ad Hoc Communications and Networks Workshops, June, 2008.

\bibitem{bahrak:en:lejp:11}
L. Lai, H. ElGamal, H. Jiang, and V. Poor, "Cognitive Medium Access: Exploration, Exploitation and Competition," IEEE Transaction on Mobile Computing, 2011.

\bibitem{bahrak:en:lm:05}
D. Lowd and C. Meek. ``Adversarial learning," In
Proceedings of the Eleventh ACM SIGKDD International Conference on Knowledge Discovery and Data Mining, pages 641-647, 2005.

\bibitem{bahrak:en:bns:06}
M. Barreno, B. Nelson, R. Sears, A. Joseph, and J. Tygar. ``Can machine learning be secure?," In ACM Symposium on Information, Computer and Communication Security, pages 16-25, 2006.

\bibitem{bahrak:en:nbc:08}
B. Nelson, M. Barreno, F. J. Chi, A. D. Joseph, B. I. Rubinstein, U. Saini, C. Sutton, J. Tygar, and K. Xia. ``Exploiting machine learning to subvert your spam filter," In Proceedings of the First USENIX Workshop on Large- Scale Exploits and Emergent Threats, April 2008.

\bibitem{bahrak:en:bnj:08}
M. Barreno, B. Nelson, A. D. Joseph, and D. Tygar, ``The security of machine learning," Machine Learning Journal (MLJ) Special Issue on Machine Learning in Adversarial Environments, 2008.

\bibitem{bahrak:en:tgmps:11}
K. Thomas, C. Grier, J. Ma, V. Paxson, and D. Song, "Design and Evaluation of a Real-Time URL Spam Filtering Service," IEEE Symposium on Security and Privacy 2011.

\bibitem{bahrak:en:g:60}
E.N. Gilbert, ``Capacity of burst-noise channels," Bell Syst. Tech. J., vol. 39, pp. 1253-1265, Sept. 1960.

\bibitem{bahrak:en:ss:71}
R. Smallwood and E. Sondik, ``The optimal control of partially observable Markov processes over a finite horizon," Operations Research, pp. 1071-1088, 1971.

\bibitem{bahrak:en:sb:98}
R. S. Sutton and A. G. Bareto, ``Reinforcement Learning: An Introduction," MIT press, 1998.

\bibitem{bahrak:en:w:47}
A. Wald, ``Sequential Analysis." Wiley, 1947.

\bibitem{bahrak:en:bv:04}
S. Boyd and L. Vandenberghe, ``Convex Optimization," Cambridge University Press, 2004.

\bibitem{bahrak:en:bz:07}
N. Baldo and M. Zorzi, ``Learning and adaptation in cognitive radios using neural networks," In: 1st IEEE workshop on cognitive radio networks (in conjunction with IEEE CCNC 2008), 12 January 2008, Las Vegas, Nevada, USA.

\bibitem{bahrak:en:nc:09}
T. Newman and T. Clancy, ``Security threats to cognitive radio signal classifiers," in Virginia Tech Wireless Personal Communications Symposium, June 2009.

\bibitem{bahrak:en:ckn:11}
T. Clancy, A. Khawar, and T. Newman, "Robust Signal Classification Using Unsupervised Learning," IEEE Transaction on Wireless Communication, April 2011.

\bibitem{bahrak:en:lkp:10}
M. Li, I. Koutsopoulos, R. Poovendran, ``Optimal Jamming Attack Strategies and Network Defense Policies in Wireless Sensor Networks," IEEE Transactions on Mobile Computing, vol. 9, no. 8, pp. 1119-1133, Apr. 2010

\bibitem{bahrak:en:om:93}
M. Ohya, and M. Mizu,  ``An information-theoretical approach and its application to optimal problems," Electronics and Communication in Japan, Part 3, Vol 76, No. 6, 1993.

\end{thebibliography}
\end{document}